\documentclass[aps,prl,twocolumn,groupedaddress,superscriptaddress]{revtex4-2}

\usepackage{amssymb,mathrsfs}
\usepackage{placeins}
\usepackage{bm}
\usepackage{amsmath}
\usepackage{enumerate,algorithmicx,algorithm}
\usepackage{algpseudocode}
\usepackage{physics}
\usepackage{comment}
\usepackage{xcolor}
\usepackage{graphicx}
\usepackage{array,multirow}
\usepackage[utf8]{inputenc}
\usepackage[T1]{fontenc}
\usepackage{float}
\usepackage{dsfont}
\usepackage[newcommands]{ragged2e}
\usepackage{siunitx}
\usepackage{subcaption}
\usepackage[many]{tcolorbox} 
\usepackage[]{}
\usepackage[dvipsnames]{xcolor}

\usepackage{xr}

\makeatletter
\newcommand*{\addFileDependency}[1]{%
	\typeout{(#1)}%
	\@addtofilelist{#1}%
	\IfFileExists{#1}{}{\typeout{No file #1.}}%
}
\makeatother

\newcommand*{\myexternaldocument}[1]{%
	\externaldocument{#1}%
	\addFileDependency{#1.tex}%
	\addFileDependency{#1.aux}%
}

\makeatletter

\newcommand{\Rmnum}[1]{\expandafter\@slowromancap\romannumeral #1@}
\newcommand{\colorcaption}[2][]{
	\begingroup
	\renewcommand{\@caption@fignum@sep}{ (color online). }
	\caption[#1]{#2}
	\endgroup
}
\newcommand{\beq}{\begin{equation}}
\newcommand{\eeq}{\end{equation}}
\newcommand{\bse}{\begin{subequations}}
\newcommand{\ese}{\end{subequations}}
\newcommand{\bea}{\begin{eqnarray}}
\newcommand{\eea}{\end{eqnarray}}
\newcommand{\bem}{\begin{displaymath}}
\newcommand{\eem}{\end{displaymath}}
\newcommand{\bmat}{\begin{bmatrix}}
\newcommand{\emat}{\end{bmatrix}}

\newcommand{\bs}{\boldsymbol}
\newcommand{\bc}{\begin{center}}
\newcommand{\ec}{\end{center}}
\newcommand{\bmk}{\bm{k}}

\newcommand{\bC}{\left[C\right]}
\newcommand{\bCs}{\left[C^{*}\right]}
\newcommand{\mkc}{k_{\mathbb{C}}}
\newcommand{\mKc}{K_{\mathbb{C}}}
\newcommand{\bmkc}{\bar{k}_{\mathbb{C}}}

\myexternaldocument{SINKSG}

\graphicspath{{Figures/}}
\usepackage{hyperref}
\usepackage{flushend}

\definecolor{clr1}{rgb}{0,0,0}
\definecolor{clr2}{rgb}{0.0,0.35,0.0}

\newcommand{\figref}[1]{FIG.~\ref{#1}}

\newcommand{\supsecref}[1]{SEC.~\ref{#1} of \cite{supp}}

\makeatletter
\myexternaldocument{SINKSG}

\usepackage{hyperref}
\usepackage{flushend}

\begin{document}

	\title{Zero modes of non-abelian Dirac operator in topologically non-trivial band insulator}

	\author{Neha Kumari}
	\affiliation{Department of Physics,
		Indian Institute of Technology Delhi,
		Hauz Khas, New Delhi 110016}
	\author{Sankalpa Ghosh}
	\email[Contact author: ]{sankalpa@physics.iitd.ac.in}
	\affiliation{Department of Physics,
		Indian Institute of Technology Delhi,
		Hauz Khas, New Delhi 110016}
	
        \date{\today}
        \begin{abstract}
          We show that the local gauge-invariance of the quantum geometric tensor (QGT) defined in the two-dimensional Block-momentum (BM) space of a generic $N$-band (arising out of sub-lattice degrees of freedom) insulator represented by a tight-binding hamiltonian, leads to the existence of zero modes of non-abelian Dirac operator in such BM space. These zero modes in the two-dimensional Brillouin zone torus $T^{2}$, determines the probability amplitudes associated with the $N$-component ground state wave-function under adiabatic approximation. We analytically derive the expansion coefficients of maximally localized Wannier function in one-dimension from the solutions of these zero mode equations, and obtain momentum-space generalization of celebrated Laughlin gauge condition as a by-product.The commutator of these non-abelian Dirac operators define the algebra of the guiding center co-ordinates which provide a more general frame-work to study such systems than one based on the Landau-level analogues. In the ideal quantum geometric limit, these zero-mode solutions become self-dual momentum-space instantons, and can be linked to the non-abelian generalization of the vortexability criterion of Chern bands in momentum-space. We derive some exact solutions of such zero-mode on the complex torus geometry, and discuss implications of this framework for realistic systems such as two-dimensional topological insulators, rhombohedral $N_{L}$-layer graphene.   
        \end{abstract}
\maketitle
The spectral properties of Dirac operators \cite{esposito1998dirac} played an important role
in mathematics and physics  as evidenced through role of the celebrated 
Index theorem \cite{AtiyahSinger1963, AtiyahPatodiSinger1975} in gauge theory and topological condensed matter systems \cite{JackiwRebbi1976,AharonovCasher1979,JACKIW1981681,Lozano1989,Hatsugai1993,Semenoff2007,Witten2016,Si2017,Iugov2025}. In this Letter we show 
that the zero-modes of the non-abelian Dirac operator (NADO) determine the quantum geometry \cite{Resta2011} and associated topological features of the Brillouin zone (BZ) of a general two-dimensional (2D) lattice system with sub-lattice/orbital indices. Landau levels (LL) that play a central role in the theory of quantum Hall systems, and thereby in two-dimensional topological condensed matter systems, is cast as a special case of this more general mathematical framework. 
A set of equations was derived to establish this claim, their consequences discussed, and analytical solutions in some special cases are provided.  

The quantum geometric tensor (QGT), a metric that quantifies the quantum geometry of a given band with sub-lattice indices \cite{roy2014band,Wang2021,wang2023origin} (see \supsecref{BlochWannierIa}, \supsecref{BlochWannierIb} for details), is known to capture the complexity of the flat-band physics 
\cite{cano2026idealquantumgeometryfractional} and strong correlation \cite{Yu2025}  in a variety of 
recently discovered  strongly correlated topological phases in a host of layered materials: 
such as fractional Chern insulators (FCI) \cite{Xie2021,Park2023, Cai2023,Zeng2023,PhysRevX.13.031037}, and anomalous Hall Wigner crystal state \cite{Su2025,Devakul2024,LiangFu2024} in twisted van der Waals (vdW) heterostructures, FCI states \cite{Xie2025}, and  chiral superconductivity \cite{Han2025,qin2026extremeanisotropymetallicsuperconducting} in rhombohedral graphene superlattices. Majority of the existing theoretical frameworks devoted to the study of quantum geometry of such two-dimensional lattice systems with orbital indices  
are based on the generalization of the physics of lowest Landau levels (LLL)  \cite{Neupert2011,Regnault2011,KaiSun2011,Bernevig2012Wanierfunction,Carusotto2014Artificalmagfield,Milovanovic2014Cheninsulator,Carusotto2016Momentumspace,Devereaux2015Positionmomentumduality,Lee2017,estienne2023ideal,Mera2021geometry,Ledwith2023,Mera2024Landaulevel} in  fractional quantum Hall (FQH) systems to retrieve the Girvin-MacDonald-Platzman (GMP) algebra \cite{GMP1985,Parameswaran2012,Simon2025} of the projected 
density operators. 
Careful comparison with the recent experiments showed that this approach has limited success \cite{Shiva2024}, necessitating serious revisit. 

In this Letter we establish the following results in that direction. We begin by considering a 2D lattice system described by a general tight-binding hamiltonian with $N$ sub-lattice indices (usually called $N$-band insulator in literature). In this $N$-dimensional Hilbert space spanned by the eigenstates of this hamiltonian, we first show that the local gauge-invariance of the QGT, leads to the existence of zero modes of NADO in the space of Bloch-momentum (BM), which under adiabatic approximation completely determines the probability amplitudes associated with the $N$-component ground-state wave-function upto a gauge factor. Remarkably the solutions of these zero-mode equations of NADO under suitable gauge choices, allow us to analytically retrieve the expansion co-efficients of well known one-dimensional maximally localized Wannier functions (ODMLWF) in the BM-basis \cite{Resta1999,Marzari2007,Vanderbilt2011,Bernevig2011berryconnection,Liang2011Wannier}, whereas the complex phase of this expansion coefficients are shown to obey a general form of celebrated Laughlin gauge condition \cite{Laughlin1981} in BM space. 
Then we show that at the lower bound of the well known trace  condition \cite{roy2014band, Milovanovic2014Cheninsulator} satisfied by the QGT, the zero mode equations of  NADO define the generalized condition of vortexability of Chern band, and their solutions are self-dual momentum-space instantons. Subsequently, from the fact that NADO obeys the usual commutator of Yang Mills \cite{WuYang1975,Aharonov2005Chapter4} field but in the 2D BM space, the Landau level (LL) algebra can now be re-interpreted as a real-space version of the same in the presence of a uniform abelian gauge (magnetic) field. It also directly implies that the underlying space is non-commutative, but with 
non-abelian connections, which now determine the guiding center co-ordinates. Their commutator determines related algebra.  Holomorphic LLL space can be seen as a real-space realization of this more general mathematical framework. Finally, we provide some exact solutions of these self-dual equations on the complex Torus of the BZ in the ideal quantum geometric limit, which might help extending this theory further through  numerical computations in realistic systems. 

We consider a general tight binding hamiltonian to describe a $N$-band insulator in the BM-basis representation as \cite{Marzari2007,Resta2011,Liang2011Wannier,Bernevig2011berryconnection,Neupart2013,RahulRoy2015,Lee2017, Simon2020latticegeometry}
\beq H = \sum_{\bs{k},\alpha,\beta} h ^{\alpha \beta}(\bs{k})\ket{\alpha, \bs{k}}\bra{\beta, \bs{k}}. \label{Nbandham1} \eeq 
Here $\bs{k}=\{k_{x},k_{y} \}$ is the two-component parameter restricted in the first Brillouin zone (BZ), and $\alpha,\beta=1,\cdots N$ are sub-lattice or orbital (or pseudo-spin/flavor) index (Details in 
\supsecref{BlochWannierIb}).
Upon diagonalization, the hamiltonian \eqref{Nbandham1}can be written as 
\beq H_{\bs{K}}= \sum_{\bs{k},n=1}^{N}E_{n}(\bs{k})\ket{u_{n}(\bs{k})}\bra{u_{n}(\bs{k})} \label{Nbandham2} \eeq 
The orbital basis states in \eqref{Nbandham1} and the eigenbasis in \eqref{Nbandham2} are related to each other through an $U(N)$ transformation ( see \eqref{QGTUN} in \supsecref{QGTensor2}).The relation between these basis states with the orbital basis state (detailed in \eqref{TBorbit5} of \supsecref{BlochWannierIb} and associated discussion).
\bse \label{QGTnotation1}
 \bea \ket{u_{n}(\bs{k})} & = & \sum_{\alpha=1}^{N} u_{n \alpha}(\bs{k})\ket{\alpha, \bs{k}} \\
 \ket{\alpha,\bm{k}} & = &  \frac{1}{\sqrt{N}}\sum_{\bm{R}} e^{i \bm{k}.(\bm{R} +  \tau_{a})} \ket{\alpha, \bm{R}}. \eea 
 \ese 
The eigen-basis \eqref{QGTnotation1} of the hamiltonian  defines an $N$-dimensional Hilbert space $\mathbb{C}^{N}$ \cite{Wilczek1984GaugeStructure, Bernevig2011berryconnection, Ming2010NonabelianQGT, Neupart2013}. If the spectrum contains a degenerate point \cite{WuYang1975,NIELSEN198120, NIELSEN1981173, monopole2019}, then any two vectors from $\mathbb{C}^{N}$ can be related to each other by a non-vanishing complex number leading to a complex projective space $CP^{N-1}$ \cite{Ryu_2010,Jian2013PSS,Lee2017,Mera2024Landaulevel}.

To further discuss the geometry of this Hilbert space, following Wootters\cite{Wootters1981, Bengtsson2002} we introduce
the statistical distance between two pure quantum states as 
\beq\label{fs1}
	D = \cos^{-1}(\frac{|\bra{\Psi_1}\ket{\Psi_2}|^2}{\bra{\Psi_1}\ket{\Psi_1} \bra{\Psi_2}\ket{\Psi_2}})^{\frac{1}{2}}.
\eeq
If the state in \eqref{fs1}, depends smoothly on a set of parameters, the corresponding infinitesimal distance in the parameter space can be expressed in terms of a metric which is called quantum geometric tensor (QGT)(details in \supsecref{QGTensor1}) which should be invariant under local gauge transformation \cite{Provost1980} (A full derivation is given in  \supsecref{QGTensor1}). 

Now, consider an adiabatically evolving  general ground state defined in this $N$ -dimensional Hilbert space of \eqref{Nbandham2} spanned by the basis vectors given in \eqref{QGTnotation1}(a),
in the two-dimensional parameter space $\bs{k}=\{k_{x},k_{y}\}$ defined in first BZ (FBZ), and can be written as  
\beq \ket{\Psi(\bs{k})}=\sum_{n=1}^{N}c_{n}(\bs{k})\ket{u_{n}(\bs{k})}, \text{with}\sum_{n=1}^{N}\abs{c_{n}}^{2}=1 \label{QGTstate} \eeq
such that the distance element between two-such state-vectors differing from each other by infinitesimal 
separation in parameter space can be written in terms of QGT as  
\bse \label{metric}
    \bea  \dd{s}^{2} & = & \abs{\ket{\Psi(\bs{k}+\dd\bs{k})}-\ket{\Psi(\bs{k})}}^{2} \\
& = & \braket{\partial_{\mu}\Psi(\bs{k})}{\partial_{\nu}\Psi(\bs{k})}dk^{\mu}dk^{\nu} \\
& = & \bCs^{T} g_{\mu\nu} \bC dk^{\mu}dk^{\nu}. \eea
\ese 
Here, $\partial_{\mu}=\frac{\partial}{\partial k_{u}}$, $g_{\mu \nu}(\bmk)$ is the parameter dependent $N_{D} \times N_{D}$ quantum geometric tensor (QGT) whose $N \times N$ matrix-elements in the sub-lattice space are 
\bse
\bea g_{\mu\nu}^{mn} & = & \bra{\partial_{\mu}u_{n}(\bs{k})}
[\mathds{1}-P(\bs{k})]\ket{\partial_{\nu}u_{m}(\bs{k})} \label{QGMatrixelement}\\
& = & \Gamma^{\text{FS}}_{\mu \nu}-\frac{i}{2}F_{\mu \nu} \label{FSBC} \eea
\ese
Here, $^{T}$ stands for transpose, and $\bC^{T}=\begin{bmatrix}c_{1}(\bs{k}) & 
c_{2}(\bs{k}) & \vdots & c_{N}(\bs{k}) \end{bmatrix}$ is a $N \times 1$ row vector,
and $\Gamma^{\text{FS}}_{\mu \nu}$ and $F_{\mu \nu}$ in \eqref{FSBC} are respectively  
Fubiny-Study (FS) metric, and (non-abelian) Berry curvature (BC)(see \supsecref{QGTensor4c}). It is to be noted $N_{D}=2$, and $N$ refer to the dimension of parameter space, and orbital space respectively. 
To impose the local-gauge invariance condition on QGT \cite{cheng2013quantumgeometrictensorfubinistudy}, we write (details in \supsecref{QGTensor2})
\beq \ket{\partial_{\mu} \Psi(\bs{k})}= \ket{D_{\mu}\Psi(\bs{k})}+ [\sum_{n=1}^{N}c_{n}(\bs{k})\cdot[1-P(\bs{k})]\cdot \ket{\partial_{\mu}(u_{n}(\bs{k})}] \label{QGTM1} \eeq 
with the projection operator (PI) (details in \eqref{QGTprojection} in \supsecref{QGTensor2}) 
\beq P(\bs{k}) = \bigoplus \sum_{n=1}^{N} \ket{u_{n}(\bs{k})}\bra{u_{n}(\bs{k})} \label{projection1}
\eeq 
whose action on the state $\ket{\partial_{\mu}u_{n}(\bs{k})}$
\cite{Ming2010NonabelianQGT,Dalibard2011gaugepotential,Neupart2013,MeraMits2022,Yu2025} generates a 
a non-abelian Berry connection or Wilczek-Zee (WZ) connection \cite{WuYang1975,Wilczek1984GaugeStructure}, 
defined as 
\beq A_{\mu}^{mn}(\bs{k})= i \braket{u_{m}(\bs{k})}{\partial_{\mu}u_{n}(\bs{k})}\label{WilZee} \eeq
since (see \eqref{Wilzee1} in \supsecref{QGTensor2}) for details) $P(\bs{k}).\ket{\partial_{\mu}u_{n}(\bs{k})}=
-i \sum_{m=1}^{N}A_{\mu}^{mn}\ket{u_{m}(\bs{k})}$. 
Under quantum adiabatic limit, the local gauge invarinace implies parallel transport condition \cite{Bradlyn2022,cano2026idealquantumgeometryfractional} in the sense of Levi-Civita. This is enforced by demanding(\eqref{QGTnotation5} in \supsecref{QGTensor2} and \eqref{NbyNWilzee1} in \supsecref{QGTensor3}),
\bse \label{QGTparallal1}
\bea \ket{D_{\mu}\Psi(\bs{k})} & = & 0 \\
\Rightarrow (\partial_{\mu} \mathbb{I} - i \bs{A}_{\mu}(\bs{k}))\bC & = & 0 \eea 
\ese 
In the above equation we introduce non-abelian Dirac operator (NADO) in Bloch momentum space
(Details in \eqref{NbyNWilzee2} with \eqref{NbyNWilzee1}, and \eqref{QGTparallel} in \supsecref{QGTensor2}), appearing 
as $D_{\mu}(\bs{k}) = \partial_{\mu} - i A_{\mu}(\bm{k})$, and  matrix elements of $A_{\mu}$ are given in \eqref{WilZee}. The solutions of equation \eqref{QGTparallal1} are zero modes of NADO, 
that are derived for a generic $N$-band tight binding hamiltonian given in \eqref{Nbandham1} after demanding local gauge-invariance of the QGT. Therefore these zero-mode solutions completely determine the probability amplitudes of a general ground state wavefunction \eqref{QGTstate} in this $N$-dimensional Hilbert space upto a gauge transformation.

 The  matrix elements of $\Gamma_{\mu \nu}^{\text{FS}}(\bs{k}),F_{\mu \nu}(\bs{k})$ in \eqref{FSBC} can be written in terms of matrix elements in \eqref{QGMatrixelement}( see \eqref{FSB} in \supsecref{QGTensor4b}).
Since $(\mathds{1}-P)^{2}=(\mathds{1}-P)$, \eqref{QGMatrixelement}
can be interpreted as inner-product of two vectors:
\bse \label{CSinequality1}
\bea \ket{\alpha_{\mu}^{m}} & = & (\mathds{1}- P(\bs{k})) \ket{ \partial_{\mu}u_{m}(\bs{k})} \\
\ket{\alpha_{\nu}^{m}} & = & (\mathds{1}- P(\bs{k})) \ket{ \partial_{\nu}u_{m}( \bs{k})}.\eea
\ese
Applying Cauchy-Schwarz inequality to the matrix elements of the QGT defined in \eqref{QGMatrixelement} one can readily obtain the well known inequality\cite{roy2014band, Milovanovic2014Cheninsulator,Peotta2015,Ozawa2021relation,Yu2025}(details in \supsecref{QGTensor3}) 
\beq\label{trace1}
	\sqrt{\det(\Gamma^{\text{FS}}(\bs{k}))}  \ge \frac{|F_{12}(\bs{k})|}{2}.
\eeq
In a two-dimensional parameter space and due to the positive semidefinite nature of QGT, \eqref{trace1} implies trace-condition 
\beq \label{trace2}
\text{Tr}(\Gamma^{FS}(\bs{k}))  \ge  |F_{12}(\bs{k})|.
\eeq 
The above trace-condition is similar as the topological bound (Bogomolny bound) on energy functional of an isotropic ferromagnet studied in the pioneering work by Belavin and Polyakov \cite{Belavin1975,Fukui2025} (details in \supsecref{QGTensor5}), and this
lower bound is called ideal quantum geometric limit in recent literature\cite{Ledwith202FCI, Kwon2024}. To progress further we define complex wave-vectors, their derivative operators as 
\bse \label{complexBZ1}
\bea k_{\mathbb{C}} &=& k_{x} +ik_{y} ~;~
     \bar{k}_{\mathbb{C}} =  k_{x} -ik_{y} \\
     \partial_{k_{x}} & = & (\partial_{k_{\mathbb{C}}}+\partial_{\bar{k}_{\mathbb{C}}}) ~;~
     \partial_{k_{y}} =  i(\partial_{k_{\mathbb{C}}}- \partial_{\bar{k}_{\mathbb{C}}}). \eea
     \ese 
Substituting the above definitions in the lower bound of the trace-condition \eqref{trace2} a straightforward algebra gives (details in \supsecref{QGTensor4c})
\beq Q(\bs{k})\ket{\partial_{\bar{k}_{\mathbb{C}}}\Psi(\bm{k})}  =  0 \label{vortexability} \eeq 
where $\mathds{1}-P=Q$. Eq. \eqref{vortexability} (see derivation of \eqref{QGTspace8e} and \eqref{Nvortexability} in \supsecref{QGTensor4c}) implies  
\beq D_{\mu}\bC = \pm i \varepsilon_{\mu \nu}D_{\nu}\bC, \label{Self-Dual} \eeq 
namely that the zero mode solutions that satisfy the ideal quantum geometric limit through \eqref{trace2}  are, momentum space instanton and anti-instanton solutions that are self-dual.

The commutators of NADOs are given by\cite{Niu2005nonabelian,Palumbo2021nonabelian, Estienne_2011nonabelian}(details in \eqref{commutator1} of \supsecref{QGtensor4a}):
\beq \comm{D_{\mu}(\bs{k})}{D_{\nu}(\bs{k})}= - i F_{\mu\nu}(\bm{k}) \label{commutator} \eeq
where
\beq F_{\mu \nu}(\bs{k})=\left[ \partial_{\mu} A_{\nu}(\bm{k}) - \partial_{\nu}A_{\mu}(\bm{k})- i \comm{A_{\mu}(\bm{k})}{A_{\nu}(\bm{k})} \right]. \label{FieldConnection} \eeq 
The position operator $\hat{x}_{\mu}$ in the Bloch momentum space is represented as $-i \partial_{\mu}$. Hence, the WZ connection in \eqref{WilZee} can be interpreted as representation of this position operator in the $N$-dimensional sublattice
space. Therefore we can re-write \eqref{commutator} as 
\bse
\bea D_{\mu}(\bs{k})  \equiv  i \hat{\bar{x}}_{\mu},&\text{with}&\hat{\bar{x}}_{\mu}(\bs{k})=-(i\partial_{\mu}+A_{\mu}(\bs{k})) \label{noncommutative1} \\
\comm{\hat{\bar{x}}_{\mu}(\bs{k})}{\hat{\bar{x}}_{\nu}(\bs{k})} & = & -F_{\mu \nu}(\bs{k}) \label{noncommutative2} \eea
\ese
Here $\hat{\bar{x}}_{\mu}(\bs{k})$ is the modified or gauged position operator due to the projection in the sublattice space, and the relation \eqref{noncommutative2} provides the non-abelian generalization of commutator of the guiding center co-ordinates \cite{Feldman1970}, but now in BM space, implying that guiding center co-ordinates in the sub-lattice space of given band are non-commutative \cite{connes2000noncommutativegeometryyear2000} with non-abelian connections. This has important consequences for GMP algebra \cite{Haldane2011,MurthyShankar2012}, and transport \cite{Peotta2015,VermaQueiroz} properties, such as anomalous Hall effects \cite{Fang2003}.
We may recall that abelian (single-component) version of \eqref{commutator} in co-ordinate space is well-known Heisenberg-Weyl algebra yielding LLs,
\beq \comm{(\bs{p}+\frac{e}{c}\bs{A})_{\mu}}{(\bs{p}+\frac{e}{c}\bs{A})_{\nu}} = -i \frac{\hbar^{2}}{\ell_{B}^{2}}. \label{HWcomm} \eeq
Here $\bs{p}, \bs{A}$ are respectively canonical momentum, the vector-potential associated with the magnetic field in real space, $\ell_{B}=\sqrt{\frac{\hbar c}{eB}}$ is the magnetic length. Eq. \eqref{HWcomm} describes non-commutative geometry of the LLs, and allows one to form a magnetic-von Neuman lattice for a constant magnetic field in the coherent-state basis \cite{Haldane1985,perelomov1986generalized,Ferrari1990,Haldane22018}. 
Eqs. \eqref{commutator}, \eqref{noncommutative2} reinterpretes \eqref{HWcomm}, 
as a co-ordinate space, abelian generalization of a more general result, with zero-modes of NADO in $\bs{k}$-space replacing the LLL condition.
Construction of analogues of magnetic von Neumann lattice and coherent state basis in Bloch-momentum space for non-abelian connection and associated curvature, however thus far turned out to be more complex \cite{Lee2017,Okuma2024Vortex,Mera2024Landaulevel,mera2025nonabelain, Fujimoto2025}.

In the preceding discussion we established that zero-modes of NADO in $\bs{k}$ space completely determines the ground state a $N$-band (sub-lattice) tight binding hamiltonian in two-spatial dimension. In general, the zero mode equations in \eqref{QGTparallal1}, or \eqref{Self-Dual} in the non-ideal, and ideal quantum geometric limit, being first order differential equations, can be solved numerically with periodic boundary condition on BZ torus for further calculation in realistic materials. However, it will be useful if at least for some special cases, analytical closed form solutions can be provided. We shall identify some such cases in the following discussion. 

{\it Exact solution for cases in the ideal quantum geometric limit:} If WZ connection in \eqref{WilZee} is diagonal, the $N$ zero mode equations in \eqref{QGTparallal1} or \eqref{vortexability}
are decoupled. Under such situation remarkably, Eq.  \eqref{vortexability}(\eqref{Nvortexability} in \supsecref{QGTensor4e}) 
admits analytical closed form solutions in the periodic Brillouin zone in terms of Jacobi theta functions \cite{WhittakerWatson1927,kharchev2015theta,Ramanujan1918Trigonometric,Chang-Young_2005,Haldane12018,Haldane22018} in some cases. We demonstrate this for the  $N=2$, which is the minimal value for $N$ to have non-trivial WZ connection. This may model some practically relevant two-band systems in the ideal quantum geometric limit, namely when lower bound of Eq. \eqref{trace1} is saturated. The matrix-valued non-abelian Wilczek-Zee connection (see \eqref{2by2Wilzee}, \eqref{2by2WilZeecomplex}, \eqref{NbyNWilzee1}) $A_{\mu}$, and the corresponding matrix valued field tensor defined in \eqref{FieldConnection} can be  written  \bse
\bea A_{\mu}(\bs{k}) & = & \sum_{i=1}^{3} g \frac{\sigma^{i}}{2i}A^{i}_{\mu}(\bs{k}) \label{SU2A} \\
F_{\mu \nu}(\bs{k}) & = &  \sum_{i=1}^{3}g \frac{\sigma^{i}}{2i}F^{i}_{\mu \nu}(\bs{k}) \label{SU2F}
\eea
\ese 
in terms of  Pauli matrices which are the three generators of the two-dimensional representation of the group $SU(2)$, with  $F_{\mu \nu}(\bs{k})$ in the expression of \eqref{SU2F} satisfies \eqref{FieldConnection}, with suitable redefinition. Particularly if the expansion  \eqref{SU2A} contains diagonal matrix elements, or a single Pauli matrix (and unit matrix), then in the relation \eqref{FieldConnection} the third term which is a commutator, vanishes. This makes the zero mode equations decoupled and solvable for $N=2$. A similar expansion for a general $SU(N)$ and consequent decoupling is more difficult\cite{Arovas1999sun}.

\begin{widetext}
\begin{figure}[H]
	\centering
	\includegraphics[width=0.8 \linewidth]{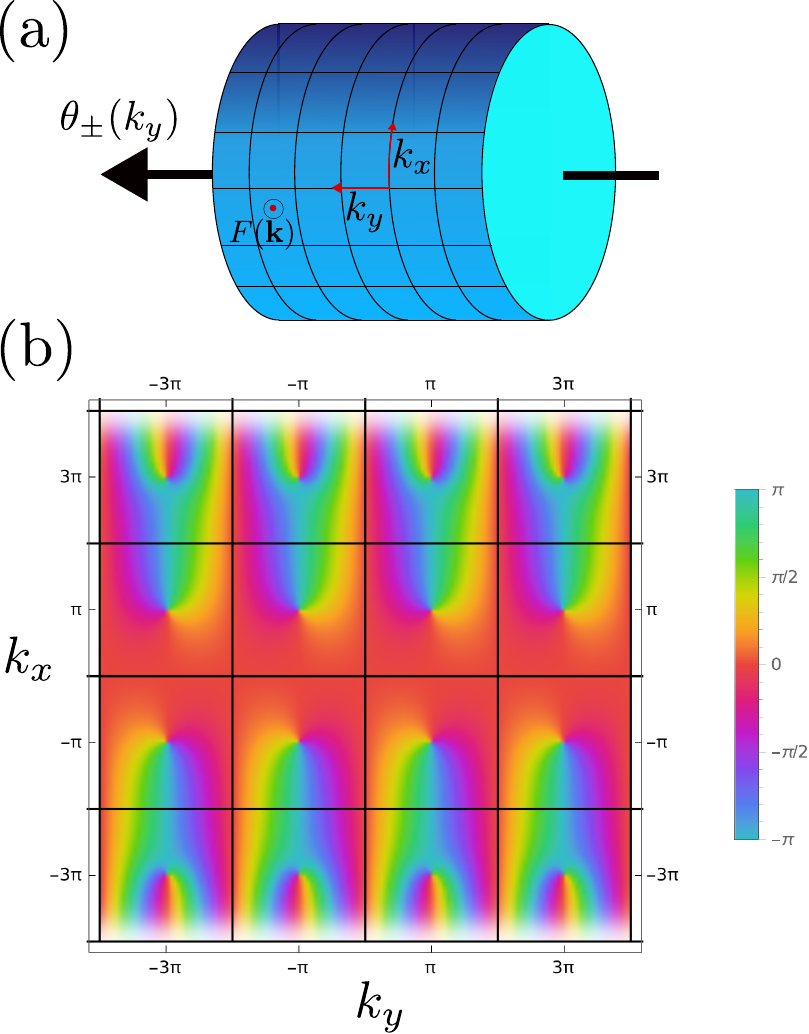}	
	\caption{ (a) A schematic depiction of Eq. (35) in the main text.(b) A plot for the Jacobi Theta function given in \eqref{Jacobithetasolution} for a square lattice in Bloch space. Location of zeoros at the center of each BZ is visible. It is plotted in Mathematica, such that both the argument and absolute value of the function is plotted to make the zeros and the winding around them visible using a color-axis.}
	\label{LaughlinTheta}
\end{figure}
\end{widetext}

Further we set the off-diagonal term in WZ connection \eqref{WilZee} to zero,
and consider \eqref{Self-Dual}(see \eqref{Nvortexability} in \supsecref{QGTensor4e}) when $\bC=\begin{bmatrix}c_{1} & c_{2}\end{bmatrix}^{T}$, when amplitudes belong to $CP^{1}$-space to get  
\beq (\partial_{\bar{k_{c}}}\mathds{1} +i A_{\bar{k_{c}}})\bC = 0 \label{CP1eq} \eeq 
where $A_{\bar{k_{c}}}$ is obtained by setting off-diagonal elements to $0$ in \eqref{2by2WilZeecomplex} in \supsecref{QGTensor4e}. 
Following \cite{AharonovCasher1979} we set  
\beq c_{n}=\exp{\Phi_{n}}f(k_{\mathbb{C}}). \label{CPN1} \eeq
As a consequence of gauge-invariance and holomorphicity on a complex torus $T^{2}$,
solutions of \eqref{CP1eq} can be determined from (details, and notations in \supsecref{Instanton1})
\bse 
\bea (\partial_{k_x} + i \partial_{k_y})f(k_{\mathbb{C}}) & =&  0. 
\label{Jacobithetaequation} \\
\bs{\nabla}^{2}_{\bs{k}}\Phi^{1,2} &=& F^{1,2}(\bs{k}) \label{Phiequation} \eea
\ese 
These solution of \eqref{Jacobithetaequation} can be written in terms of Jacobi theta-functions(details in \supsecref{Instanton2}, see \figref{LaughlinTheta}(b)) where we follow the notation of \cite{WhittakerWatson1927}, 
\beq f(k_{\mathbb{C}})=\vartheta_{3}(u|\tau)  = \sum_{n\in \mathbb{Z}} e^{\pi i \tau (n +m_2)^2} e^{2\pi i (n+m_2) (u+m_1)}. \label{Jacobithetasolution} \eeq 
In the 2D BZ, we define two reciprocal lattice vectors in complex notations as $b_{1,2}$, $\tau=\frac{b_{2}}{b_{1}}=a+ib$ such that the dimensionless complex variable 
 $u = \frac{\mkc}{b_{1}}+ m_1  + m_2 \tau$, with $m_{1,2}\in \mathbb{Z}$. Upon compactification, the BZ becomes a complex 2-Torus $T^{2}=\mathbb{C}/\Lambda$, where $\Lambda=m_{1}+m_{2}\tau$.
Expression \eqref{Jacobithetasolution} (for details see \eqref{theta4a}, \eqref{thetamultiplier} and discussion in \supsecref{Instanton2})shows the translational properties of $\vartheta_{3}$ with the multiplier which is a consequence of its satisfying the co-cycle condition \cite{Chang-Young_2005}. In the real space for an abelian magnetic field in the Landau problem, this becomes the algebra of the magnetic translation operators\cite{Zak1964_MTG1,Ferrari1990,TENJINBAYASHI2007460,TekinPopov2021} which satisfy a 2-cocycle condition. Eq. \eqref{Phiequation}, which is a Poisson equation \cite{griffiths2023introduction} with the charge density replaced by Berry curvature, 
can be also be solved with the help of Green's function on a complex 2-Torus $T^{2}$. Following \cite{lin2010elliptic, bergweiler_eremenko_2016}, for $T^{2}$ the Green's function is given as ( details in \supsecref{Instanton3})
\beq G^{(T^{2})}(\mKc)  =  -\frac{1}{2\pi}\log(\abs{\vartheta_{1}(\mKc)})+\frac{1}{2b}\bar{k}_{y}^{2}+C(\tau), \label{GreenfunctionTorus2} \eeq
where $\bar{k}_{y}= \Im{\frac{\mkc}{b_{1}}}$, and $\mKc=b_{1}u$, and $C$ is a $\tau$ dependent constant, and following \cite{WhittakerWatson1927} $\vartheta_{1}$
is defined in 
\eqref{theta13relation} of \supsecref{Instanton3}. $\Phi(\bs{k})$ can now be determined as 
\beq
	\Phi(\mkc) = \int F(\mkc^{\prime}) G(\mkc^{\prime}-\mkc)d\mkc^{\prime} d\bmkc^{\prime} \label{Phisolution}\eeq 
    where the expression for $G(\mkc^{\prime}-\mkc)$, and  its relation with \eqref{GreenfunctionTorus2} defined in \eqref{GreenTorus1c} and \eqref{GreenTorus1b} of \supsecref{Instanton3}. Thus, for a given $F_{\mu\nu}(\bs{k})$ the solutions \eqref{CPN1} of Eq. \eqref{Self-Dual} can be written in a closed form, with WZ connections \eqref{WilZee} only containing diagonal terms.

Generalization of these solutions to $N > 2$ which corresponds to $CP^{N-1}$ space will be interesting problems in future \cite{AGUADO20022, TENJINBAYASHI2007460}. It is also instructive to note the differences of these solutions with the solutions of $CP^{N-1}$ non-linear sigma problem on an unbounded plane (for a review see chapter 4 \cite{Rajaraman1982Solitons}) that were studied extensively in field theory and quantum Hall systems. 

Few comments are in order. In \eqref{CPN1} $f_{n}$ contains $0$'s of $\vartheta_{3}$ at $u=\frac{1+\tau}{2}$ for any general 2D lattice. At these zeros, both the probability amplitude in $\bC$ vanishes, leading to the attachment of vortices (\eqref{thetaf005} in \supsecref{Instanton3}). Because of translational invariance, these zeros are periodically repeated (\eqref{thetaf006} in \supsecref{Instanton3}). In general solutions of  \eqref{CPN1} may contain $N$-component vorticity to the electronic wavefunction in each BZ. For cases, when WZ connection retains off-diagonal terms, these analytical solutions may help to get solutions numerically using established techniques \cite{Sankalpa2000QH}. 

{\it Solutions in non-ideal quantum geometry:} Most topological condensed matter models do not satisfy the lower bound of the trace-formula \eqref{trace2}. Their quantum geometry can be termed as non-ideal in general. Two well known examples are two-band lattice Dirac hamiltonian (BHZ model)
introduced in the pioneering work(details are given in \supsecref{QSH})
\cite{QSHE2006,SCZ2006paper1,SCZ2006paper2, BHZ2006,SCZ2008paper3, bernevig_topological_2013},
and the effective two-orbital continuum model of rhombohedral stacked $N_{L}$ layer (RNLG) graphene\cite{HKMIN2008,pgqh-wm5v}
(subscripts refer to the name of the models)
,
\bea  
H(\bs{k})_{\text{BHZ}} & = & \varepsilon(\bs{k})\mathds{1} + \sum_{i=1}^{3}d_{i}\sigma^{i}=\varepsilon(\bs{k})\mathds{1} +\bs{d} \cdot \bs{\sigma}. \label{twolevel1} \\
H_{\text{RNLG}} & = &  
	\begin{bmatrix}
		-D & \nu_{N_{L}}(k_x-ik_y)^{N_{L}} \\
		\nu_{N_{L}} (k_x+ik_y)^{N_{L}} & D
	\end{bmatrix}. \label{RNGHamiltonian} 
	\eea 
In \eqref{twolevel1} $\bs{d}$ is a three-component vector
(see \eqref{spinhallmagnetic}, \eqref{BWZtheta}, \eqref{BWZphi} in \supsecref{QSH} for details),where in \eqref{RNGHamiltonian} $D$ denotes the displacement field energy applied in the direction perpendicular to the multilayer stack, and $\nu_{N}$ is the generalized Dirac (Fermi) velocity. Corresponding Chern numbers satisfy $C_{1}^{\text{RNLG}}=N_{L}C_{1}^{\text{BHZ}}$ (\eqref{Chern4} in \supsecref{QSH} and \eqref{Chern5} in \eqref{RNG}).
In either cases, the decoupled zero-mode equations \eqref{QGTparallal1} takes the form 
\beq [(\partial_{\mu} - iA^{0}_{\mu}(\bs{k}))\sigma^{0} -
iA^{3}_{\mu}(\bs{k})\sigma^{3})]\begin{bmatrix}c_{1} \\ c_{2} \end{bmatrix} = 0 \label{YM2} \eeq
on a two-torus $T^{2}$ in BZ. 
In the BHZ model described by hamiltonian \eqref{twolevel1}, The non-abelian Berry connection in can be recast in the form of \eqref{SU2A} by identifying ( Derivations of \eqref{YMcon4}, \eqref{YMcon4a} to \eqref{YMcon4c} in \supsecref{QSH})
\bse \label{YMcon3}
\bea A^{0}_{\mu}(\bs{k})  &=& i\partial_{k_{\mu}}\phi(\bs{k}); A^{1}_{\mu}(\bs{k}) =  A^{2}_{\mu}(\bs{k}) =  0 \\
A^{3}_{\mu}(\bs{k})  &= & -i\cos{\theta(\bs{k})} \partial_{k_{\mu}}\phi(\bs{k})
      ;~g  =  -1 \eea
     \ese 
We note that $A_{\mu}^{i}$ in \eqref{YM2}, and in \eqref{SU2A} differ from each other by a $i$ (see from \eqref{YMcon2a} to \eqref{YMcon4c} in \supsecref{QSH}). Since both the models \eqref{twolevel1} and \eqref{RNGHamiltonian} were widely studied, it will be more instructive to prove analytically a general result about the solutions of zero-mode equations of NADO \eqref{YM2} in the non-ideal quantum geometry. 
Consider solutions of the following form, 
\beq c_{n}(k_{x},k_{y})=\exp{i\Phi_{n}}f_{n}(k_{x},k_{y}), \label{NICPN1} \eeq
where $n=1,2$. The solutions can be written as  (details in \supsecref{Instanton1b} from \eqref{NIYM1} to \eqref{f1solc} and \eqref{f1sold})
\beq c_{1,2}= \exp{-i[C^{1,2}_{0}+\int_{0}^{k_{x}}d k'_{x}A_{k'_{x}}^{\pm} 
-k_{x}(\frac{\theta^{\pm}(k_{y})}{2\pi}+ m )] }. \label{NIsolCP1} \eeq
where $A^{\pm}_{k_{x,y}}(\bs{k})=A^{0}_{k_{x,y}}(\bs{k}) \pm A^{3}_{k_{x,y}} (\bs{k})$ with $A_{\mu}$ is given in \eqref{WilZee}.
The solutions in \eqref{NIsolCP1} can be identified with the expansion coefficients of ODMLWF 
given in \cite{Bernevig2011berryconnection,Liang2011Wannier} upto a gauge factor. Here the Berry or WZ flux 
$\theta^{\pm}(k_{y})=\int_{0}^{2\pi}dk'_{x}A_{k'_{x}}^{\pm}$ defined in \eqref{1DBerryP} and \eqref{1DBerryM} in \supsecref{Instanton1b}, and $A_{k_{y}}^{\pm}=0$. The periodic boundary condition on the phase $\Phi_{1,2}$ (see
\figref{LaughlinTheta} (a)),
\beq \Phi_{1,2}(k_{x} + 2\pi, k_{y})-\Phi_{1,2}(k_{x},k_{y}) = \frac{\theta^{\pm}(k_{y})+2m \pi}{2} \label{Laughlingauge2} \eeq is the generalization of celebrated Laughlin gauge argument \cite{Laughlin1981} for the case of the lattice system. 

In summary, this Letter presents the zero-modes of the NADO as the general mathematical framework to understand the topological properties of band structures of 2D layered materials with sub-lattice degrees of freedom, a consequence of the local gauge invaraince of QGT. These zero-modes determine the ground state wave-functions of such systems both in the non-ideal and ideal quantum geometric limit. In the first case, the solutions reproduce the probability amplitudes associated with one-dimensional maximally localized Wannier functions for a topologically non-trivial $N$-band insulator, with the BM space version of celebrated Laughlin gauge condition as a by-product. 
In the second case solutions are self-dual vortices that are momentum space instantions on a complex torus, leading to generalized vortexability criterion for Chern band. This framework consequently establishes that the underlying space in such topologically non-trivial lattice geometry, and corresponding guiding center co-ordinates are non-commutative with non-abelian gauge connection, thereby replacing the holomorphicity of LLL structures by a more generalized criterion. We augment our theoretical propositions with some exact closed form solutions for such zero modes that can help more general numerical computations in future. Particularly understanding the effect of interaction using this zero-mode solutions, comparing these results with recent experiments, and forming a related  effective low-energy long-wavelength theory for topologically non-trivial strongly correlated phases in such layered material will be interesting future directions to pursue. Pioneering work in Quantum Hall systems \cite{ODLRO1987}, fermion-vortex problems in  non-linear field theory \cite{Vega1978, JuliaZee1975, Lozano1989} may provide the necessary stimulus.

\textbf{Acknowledgement:}NK is supported by a fellowship from MOE, Govt. of India. SG acknowledges helpful discussions with Patrick Ledwith, Shaffique Adams and Victor Galitski during a visit to Aspen Center for Physics (ACP), and with Allan H. Macdonald in GRC conference in USA after the first draft of the paper was posted in arXiv. These visits were funded by IIT Delhi PDF grant PD16125, and ACP visit was partially supported by National Science
Foundation (NSF) grant PHY-2210452.

\bibliography{reference7}
\vskip 2cm 

\end{document}